\documentclass[preprint,12pt,amsmath,amssymb]{aastex}

\usepackage{graphicx}
\usepackage{dcolumn}
\usepackage{epsfig}

\begin{document}

\title{Inhomogeneity in the  Supernova Remnant Distribution as the Origin of the PAMELA Anomaly}

\author{Nir J. Shaviv$^{1}$, Ehud Nakar$^2$ \& Tsvi Piran$^{1}$}

\affil{
1. Racah Institute of Physics, Hebrew University of Jerusalem, Jerusalem 91904, Israel \\
2. The Raymond and Beverly Sackler School of Physics \& Astronomy,
Tel-Aviv University, Tel-Aviv 69978, Israel}


\begin{abstract}
Recent measurements of the positron/electron ratio in the cosmic ray
(CR) flux exhibits an apparent anomaly \citep{Adriani}, whereby this
ratio increases between 10 and 100 GeV. We show that inhomogeneity
of CR sources on a scale of order a kpc, can naturally explain this
anomaly. If the nearest major CR source is about a kpc away, then
low energy electrons ($\sim 1$~GeV) can easily reach us.  At higher
energies ($\gtrsim 10$~GeV), the source electrons cool via
synchrotron and inverse-Compton  before reaching Earth. Pairs formed
in the local vicinity through the proton/ISM interactions can reach
Earth also at high energies, thus increasing the positron/electron
ratio. A natural origin of source inhomogeneity is the strong
concentration of supernovae in the galactic spiral arms. Assuming
supernova remnants (SNRs) as the sole primary source of CRs, and
taking into account their concentration near the galactic spiral
arms, we consistently recover the observed positron fraction between
1 and 100 GeV. ATIC's  \citep{ATIC} electron excess at $\sim
600$~GeV is explained, in this picture,  as the contribution of a
few known nearby SNRs.  The apparent coincident similarity between
the cooling time of electrons at $10$ GeV (where the
positron/electron ratio upturn), $\sim 10$ Myr, and the CRs protons
cosmogenic age at the same energy is  predicted by this model.
\end{abstract}


PAMELA \citep{Adriani} discovered that  the CR positron/electron
ratio increases with energy above $\sim10$~GeV. This ratio should
decrease according to the standard scenario, in which CR positrons
are secondaries  formed by interactions between the primary CR
protons and the interstellar medium (ISM) \citep{Strong1998leptons}.
This apparent discrepancy is now commonly known as the ``PAMELA
anomaly''. It is commonly interpreted as evidence for a new source
of primary CR positrons, most likely WIMPs
\citep{WIMPSUSY,WIMPDecay} or pulsars
\citep{Harding,Chi,Atoyan,Hooper,Yuskel,Profumo}. ATIC \citep{ATIC}
shows an excess of CR electrons at energies of $300-800$~GeV. At
even higher energies ($1-4$~TeV) HESS measures \citep{HESS} a sharp
decay in the electron spectrum. ATIC's results are usually
considered as support for  a dark matter interpretation for the
PAMELA anomaly, where the observed excess corresponds to the WIMP
mass.

In the standard picture, CRs below the knee are thought to originate
in SNR shocks. This is indicated by synchrotron \citep{Koyama} and
inverse-Compton  \citep{Tanimori} emission of high energy electrons
in SNRs, and the $\gamma$-ray emission, which is possibly from high
energy protons \citep{Aharonian}. Theoretical models for the CR flux
describe CR propagation in the Galaxy. CRs diffuse within the disk,
and escape once they reach the halo height, $l_H \sim 1$~kpc, above
the disk. Most CR diffusion models approximate the diffusion
coefficient as $D = D_0 (E/E_0)^{\beta}$ and assume that CRs are
produced with a power-law spectrum, $N_E \equiv dN /dE \propto
E^{-\alpha}$.  The observed spectrum  is then a convolution of the
source spectrum and propagation losses, giving for the primary
electrons $\phi^-(E)\propto E^{-(\alpha_e+\beta)}$. Positrons are
secondary CRs formed from CR protons, and suffer additional
propagation loses, implying $\phi^+(E) \propto \phi_p (E) E^{-\beta}
\propto E^{-(\alpha_p+2\beta)}$, where $\phi^{\pm}$ and $\phi_p$ are
the CR positrons, electrons and protons observed fluxes. The
predicted flux ratio is $\phi^+/(\phi^- + \phi^+) \approx
\phi^+/\phi^- \propto E^{\alpha_e-\alpha_p-\beta}$, where $\alpha_e$
and $\alpha_p$ are the source power-law indices of electrons and
protons respectively. Both electrons and protons are expected
\citep{BlandfordEichler87} to have similar spectral slopes, i.e.,
$\alpha_e \approx \alpha_p$, which is somewhat larger than 2. This
is also supported by synchrotron radiation observed from SNRs, which
confirms the slope for the electrons \citep{Duric1995}.
Consequently, $\alpha_p-\alpha_e < \beta \approx 0.3-0.6$ and the
standard model predicts, in contrast to PAMELA observations, a
decreasing $\phi^+/\phi^-$.

The diffusing electrons and positrons cool via synchrotron and
inverse-Compton scattering, with ${d E / dt } = -b E^2$. This
steepens both the electron and positron spectra at an energy where
the cooling time equals the typical electron and positron age.
However, since both suffer the same  loses, this  does not affect
$\phi^+/\phi^-$. Additional effects such as  spallation and
annihilation  can be safely ignored at the energies of interest.

This standard model assumes a homogenous, or at least a smoothly
varying (on a galactic scale), source distribution
\citep{Strong1998leptons,Strong1998nucleons}. However, since in
spiral galaxies star formation is concentrated in spiral arms
\citep{LaceyDuric,ShavivNewAstronomy} and SNRs are the canonical
sources of CRs, one should consider the effect of  inhomogeneities
in the CR source distribution on intermediate scales (i.e., scales
smaller than the Galactic size but large enough such that discrete
sources do not have a strong effect) on the CR spectrum. This
inhomogeneity of sources influences the electrons/positrons spectra
via cooling which sets a typical distance scale that an
electron/positron with a given energy can diffuse away from its
source. For a homogenous distribution cooling affects  the spectra
of (primary) electrons and (secondary) positrons in the same way and
their ratio is unaffected. On the other hand, primary electrons will
be strongly affected by an inhomogeneous source distribution at
energies for which the diffusion time is longer than the cooling
time. Protons are not affected by cooling and are therefore
distributed rather smoothly in the galaxy even if their sources are
inhomogeneous. The secondary positrons (that are produced by the
smoothly distributed protons) are only weakly affected by the
inhomogeneity of the sources. This effect would induce an observed
signature  on $\phi^+/\phi^-$, with similar properties to the one
observed by PAMELA.

Motivated by this expectation we  construct, first, a simple
analytic model for diffusion from an inhomogeneous source. Consider
a source at a distance $d$  from Earth. We model the solar
neighborhood of the galaxy as a two dimensional slab  (see fig.\ 1).
The $x$ coordinate (the Galactic plane) is infinite and the $y$
coordinate (the disk height) is finite, $l_H$. The source is at the
origin and Earth is at $(d,0)$. A CR diffuses within this slab with
a constant diffusion coefficient $D(E)$,  and it escapes once
$|y|>l_H$. The contribution of CR protons that were generated at
time $t'$ to the flux  at time $t_0$ can be approximated as
\footnote{We assume for simplicity that the diffusion is one
dimensional. This results with an exponent once integrated. Two
dimensional diffusion (from a linear spiral arm) would give a less
transparent Bessel function.}:
\begin{equation}
\phi_p(d,t') \propto \frac{1}{\sqrt{Dt}}
\exp[-(t/\tau_e)-(\tau_d/2t)],
\end{equation}
where $t \equiv t_0-t'$, $\tau_e \approx l_H^2/D$ is the typical
escape time  and $\tau_d\approx d^2/D$ is the typical diffusion time
from the source to Earth. Integration over $t$ for a steady source,
yields:
\begin{equation}\label{Eq. Phi_prot}
  \phi_p(d) \propto \frac{1}{D}{\exp\left[{-\sqrt{{2\tau_d}/{\tau_e}}}\right]} ,
\end{equation}
with a similar energy dependence (via  $D$) as for uniformly
distributed sources. The average age of an observed proton is
$a = l_H(l_H+\sqrt{2}d)/2D \approx \max\{\tau_e,(\tau_e\tau_d)^{1/2}\}$.

We approximate the cooling effect on the electron's flux as
$\phi^-(d,t') \propto \phi_p(d,t') \exp [-t/\tau_c]$, where $\tau_c$
is the typical cooling time.  Integration over $t$ reads:
\begin{equation}\label{Eq. Phi_elec}
  \phi^-(d) \propto
  \frac
  {\exp\left[-2\sqrt{{\tau_d}/{\tau_c}+{\tau_d}/{\tau_e}}\right]}{D\sqrt{1+{\tau_e}/{\tau_c}}}.
\end{equation}
If $\tau_c < \min\{\tau_d,(\tau_e \tau_d)^{1/2}\}$  the
electron flux drops exponentially with decreasing $\tau_c$,
while for larger $\tau_c$ the electron flux is proportional to
$D^{-1}$ (relative to the source's spectrum). This is
different than the case of uniformly distributed sources,
which shows a shallower break at $\tau_c \approx \tau_e$
from $\phi^- (\tau_c>\tau_e) \propto \tau_e \propto D^{-1}
\propto E^{-\beta}$ into $\phi^- (\tau_c<\tau_e) \propto \tau_c
\propto E^{-1}$, both relative to the source's spectrum.

The positron source function is approximately proportional to
$\phi_p(d)$.
As positrons and electrons have the same cooling
rate,
a source at $x'$ contributes to the positron flux at
$d$  approximately $\phi^-(x'-d)$. Therefore:
\begin{equation}\label{Eq. Phi_pos}
\phi^+(d)\propto \int_{-\infty}^\infty \phi_p(x') \phi^-(x'-d) dx'
\propto
\frac{\tau_c}{D}\left(\exp\left[-\sqrt\frac{2\tau_{d}}{\tau_e}\right]-
\frac{\exp\left[-\sqrt{\frac{2\tau_{d}}{\tau_c}+\frac{2\tau_{d}}{\tau_e}}\right]}{\sqrt{1+{\tau_e}/{\tau_c}}}\nonumber
  \right).
\end{equation}
For $\tau_c \gg \tau_e$, the energy dependence of $\phi^+$ relative
to the source spectrum, $\phi^{(s)}_p$, is $\phi^+/\phi^{(s)}_p
\propto D^{-2}\propto E^{-2 \beta}$ while for $\tau_c \ll \tau_e$,
$\phi^+/\phi^{(s)}_p \propto \tau_c/D \propto E^{-\beta - 1}$. This
behavior is similar to the one from  uniformly distributed sources.

Eqs.\ \ref{Eq. Phi_elec} and \ref{Eq. Phi_pos} show that for  a
source at a distance $d$ from Earth, a turnover in $\phi^+/\phi^-$ is
observed at $E_b$ which satisfies $\tau_c(E_b) \approx
\min\{\tau_x(E_b),(\tau_e(E_b) \tau_x(E_b))^{1/2}\}$.
$\phi^+/\phi^-$  for $E<E_b$ decreases, while it increases for $E>E_b$. At
the same time the typical age of CR protons with energy $E_b$ is $a
\sim \max\{\tau_e,(\tau_e\tau_d)^{1/2}\}$. Therefore a natural
prediction of the model is $a(E_b) \gtrsim \tau_c(E_b)$ and a
comparison of the two observables can be used as a consistency test
for the model. Moreover, over a wide range of the parameter space for which $d \gtrsim l_H$, the model predicts $a(E_b) \approx \tau_c(E_b)$
regardless of the value of the diffusion coefficient $D$.

Electrons and positrons in the ISM cool as ${d E / dt } = -b E^2$
where \citep{Kobayashi} $b\approx 1.8\times 10^{-16} \rm{GeV}^{-1}
s^{-1}$ at $1~\rm{GeV}$ (and $b \approx 1.4\times 10^{-16}
\rm{GeV}^{-1} s^{-1}$ at $1~\rm{TeV}$), implying a cooling time
$\tau_{c} = 1/(b E) \approx 17$~Myr at $E \approx 10~\rm{GeV}$.
Observational constraints on the typical proton CR age are measured
at a few 100 MeV. Typical ages obtained are $18^{+8}_{-9}$~Myr
\citep{Wieden1980}, $27^{+19}_{-9}$~Myr \citep{Lukasiak} or
$30^{+21}_{-10}$~Myr \citep{Simpson1988}. At 10\ GeV,  the age
should be smaller by a factor of $\sim 1-3$, depending on the exact
energy dependance of the diffusivity. Thus, according to the
observations $a(10 {\rm GeV}) \approx \tau_c(10 {\rm GeV}) \approx
10 {\rm Myr}$. This apparent coincidence which is explained
naturally by our model encourages us to look for a dominant CR
source at a distance of a $\sim$ kpc from earth. Indeed, the nearest
spiral arm to Earth
 is the Sagittarius-Carina arm at a distance of
$\approx 1$~kpc, which is just the distance needed to explain
PAMELA's observations.

To demonstrate  quantitatively the potential of this model to
recover the observed behavior of  $\phi^+/\phi^-$, we simulated
numerically the CR diffusion  for  a realistic spiral-arm
concentrated source distribution (see also
\citealt{ShavivNewAstronomy}). Before presenting these results we
stress that all other models explaining PAMELA  invoke a new ad hoc
source of high energy CR positrons which has  a  negligible effect
on low energy CR components. However, in our model, the PAMELA
explanation is intimately related to low and intermediate energy CR
propagation in the Galaxy. Namely, by revising the source
distribution of CRs, we affect numerous properties of $\sim$\ GeV
CRs. Given that the interpretation of observations (in particular,
isotopic ratios) used to infer model parameters (such as  $D_0$,
$\beta$ or $l_H$) depend on the complete model, one should proceed
while baring in mind that these parameters may differ in our model
from present canonical values. In this sense, the objective of this
letter is not to carry a comprehensive parameter study, fitting the
whole CR data set to an inhomogeneous source distribution model.
Instead, our goal is to demonstrate the potential of the model to
explain naturally the PAMELA anomaly. To this end we use the
simplest possible model, fixing all parameters with the exception of
the halo size, $l_H$, and the normalization of the  diffusion
coefficient, $D_0$, that we vary to fit the data.

\begin{figure}
\begin{center}
\epsfig{file=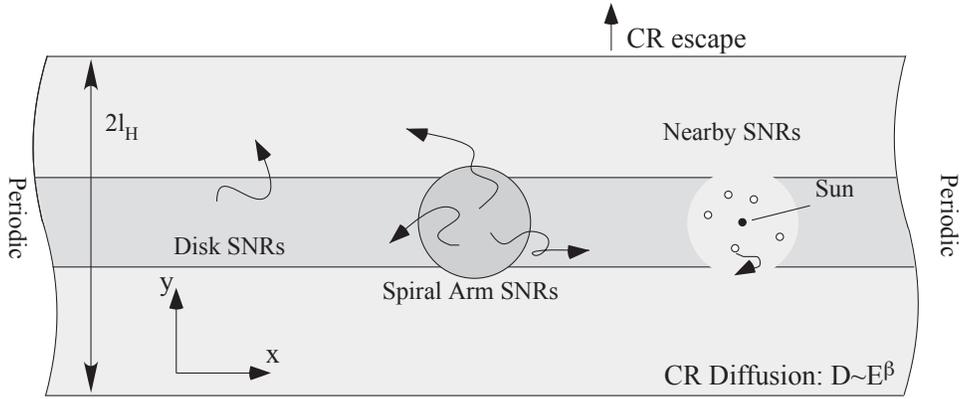,width=5in} 
\end{center}
\vskip -5mm
\caption{%
The galaxy is modeled as
a slab of width $2 l_H$, with $l_H = 1$~kpc, inside of which the CR
components diffuse. Beyond $y = \pm l_H$, the CRs escape at a
negligible time. CR sources are located in both cylinder shaped arms
with a Gaussian cross-section of width $\sigma = 300$\ pc, and disk
sources, with a vertical scale height of 100 pc. The assumption of
straight cylinders is permissible given the small spiral arm pitch
angle. This also makes the problem effectively two dimensional. We
model the Milky Way as having four spiral arms, with a pitch angle
of $i \approx 15^\circ$ \citep{Vallee}, implying that the arm
separation (in the direction perpendicular to the arm axis) is $d
\approx (\pi /2) R_\odot \sin i \approx 3$\ kpc, while the Sun is at
a distance $x \approx 1$~kpc from the nearest spiral arm. Due to the
motion of the arms, there is a small drift term carrying the CRs
away from them.  For a spiral arm periodicity of $P_s \sim 150$~Myr
\citep{ShavivNewAstronomy}, one obtains a velocity of $v_s \approx
(\pi/2)(R_\odot \sin i/P_s) \approx 20$\ km/s,  which is slower than
the two comparable diffusion times $l_H/\tau_e \approx x/\tau_x
\approx 100$~km/s. A second component resides in the disk, with an
exponential vertical decay. Because nearby sources are considered,
the The smooth disk distribution is truncated for $r<0.5$~kpc and
$t<0.5$~Myr.
  }
\label{fig:SpiralModel}
\end{figure}

The geometry of the model is described in fig.\
\ref{fig:SpiralModel}. We assume a spiral arm/disk SNe ratio of 10.
The overall normalization of the sources was fit to give the
electron spectrum at 10 GeV. The positron production was normalized
to give the positron to electron ratio at the same energy.  For the
ISM density we took the functional dependence from
\cite{Strong1998nucleons}. More on the choice of the parameters can
be found in \cite{ShavivNewAstronomy}.

We take a diffusivity of the form $D = D_0 (E/1\rm{\ GeV})^\beta$
for $E> 4\rm{\ GeV}$ and $D = D_0 (4\rm{\ GeV}/1\rm{\ GeV})^\beta$
for $E<4\rm{\ GeV}$. It was realized that such a break is required
to explain the observed break in the CR B/C ratio
\citep{Strong1998nucleons} (though it does not play an important
role here). We take $\beta = 1/3$ (corresponding to turbulence with
a Kolmogorov spectrum) and $\alpha_e = \alpha_p = 2.37$ such that
the predicted proton spectrum will be consistent with the observed
proton CR slope of 2.7. We also take $D_0=6 \times
10^{27}$~cm$^2$/sec, which reproduces the break energy in the
electron spectrum and the positron fraction. As predicted by the
analytic model the cosmogenic age we obtain in the simulation (14
Myr at 1\ GeV per nucleon) is consistent with the observations,
without fitting for it. Not surprisingly, the halo size and
diffusivity considered here are somewhat different (on the low side)
relative to standard values often found in the homogenous model.

Small scale inhomogeneities are important at energies larger than a
few hundreds GeV, for which the lifetime, and therefore propagation
distance, of electrons is so short that  the electron spectrum is
dominated by a single, or at most a few nearby sources
\citep{Atoyan,Kobayashi,Profumo}. To take this  effect into account
we truncate  the ``homogeneous'' disk component  at $r<0.5$~kpc and
age less than $t<0.5$~Myr, and we add all SNRs within this 4-volume:
Geminga, Monogem, Vela, Loop I and the Cygnus Loop,  as discrete
instantaneous sources. These sources were described using the
analytical solution \citep{Atoyan} for the diffusion and cooling
from an instantaneous point source. For the overall normalization of
the point sources, we use the synchrotron observations of SN1006,
which together with the X-rays constrain the total energy and
magnetic field \citep{Yoshida}. In particular, electrons with energy
$>1$ GeV are found to carry $\approx 2 \times 10^{48}$\ erg,
corresponding to $0.2$\% out of the total $\sim 10^{51}$~erg
mechanical energy in SNRs. We assume that all nearby sources are
similar. Note that due to their very young age, the discrete sources
contribute a negligible amount of positrons, nor do they offset the
cosmogenic age.

The lower panel of fig.\ 2 depicts $\phi^+ / (\phi^+ + \phi^-)$
obtained by the simulation. As expected from the simple analytical
model, the fraction decreases up to $\sim 10$ GeV and then it starts
increasing. This explains the so called PAMELA anomaly. As the CR
protons and antiproton spectra are  unaffected our results are
consistent with PAMELA's observations of no excess in the
anti-proton/proton ratio at the same energy range
\citep{antiprotons}. At about 100 GeV, the ratio flattens and it
decreases above this energy because of the
 injection of ``fresh" CRs from recent nearby SNRs whose high energy primary electrons don't have
time to cool.  These sources also contribute to higher energy electrons detected by ATIC.

\begin{figure}
\begin{center}
\epsfig{file=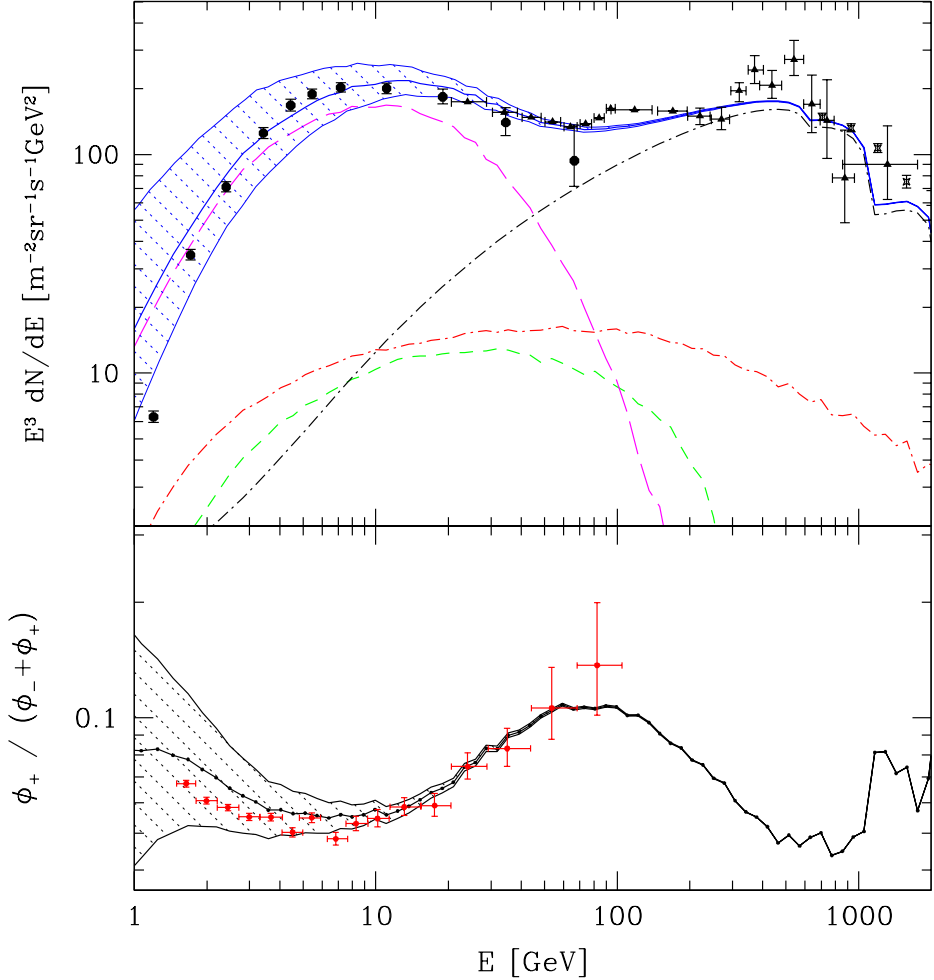,width=5.3in} 
\end{center}
\vskip -0.5cm \caption{  {\it Bottom Panel}: Model results and the
measured PAMELA points for the positron fraction.  The shaded region
is the variability expected from solar modulation effects
\citep{Clem}.
 {\it Top Panel}: The
expected electron and positron spectra -- Primary arm electrons
(long dashed purple), primary disk electrons with nearby sources
excluded (short dashed green), nearby SNRs (dot-dashed black),
secondary positrons (dot-dashed red), and their sum (blue). The
hatched region describes the solar modulation range (from 200 MV to
1200 MV). The three data sets plotted are of HEAT \citep{HEAT}
(circles), ATIC \citep{ATIC} (triangles) and HESS \citep{HESS} (open
squares). } \label{fig:Ratio}
\end{figure}

The upper panel of fig.\ 2 depicts the electron spectrum and its
constituents---primary spiral arm electrons, primary disk electrons
(without nearby sources), nearby sources and secondary pairs. There
are two bumps in the $E^3 N_E$ plot. The lower energy bump arises
from spiral arm electrons, the higher energy of which cannot reach
us due to cooling. The higher energy bump, which corresponds to the
ATIC peak, is due to a few nearby SNRs. The three ``steps" are due
to the cooling cutoffs from Geminga, Loop I and the Monogem SNRs.
Note that the high energy behavior is very sensitive to the exact
diffusion model parameters and the poorly constrained SNR energy
output in electrons. Thus, the ATIC peak is not a prediction of the
model but rather one possible outcome.

While the predictions for $\phi^+/\phi^-$ for the spiral
arms CR  model are very different than for  a homogenous sources
distribution, the effect on the electron spectrum is much more
subtle. Both models predict a break of the electron spectrum at $10$
GeV,   The break predicted by spiral arm  model is from a
power law to an exponential, while in the homogenous
model it is a broken power-law. Given that above $\sim 100$
GeV the electron spectrum is strongly affected by the sources that
produces the ATIC bump (e.g. local
SNRs), the energy range between 10 to 100 GeV is too short  to distinguish, based on the
electron spectrum alone, between the two
models. Thus, while both models can adequately reproduce the observed
electron spectrum (at least up to 100 GeV),  only the
inhomogeneous source model can explain the observed $\phi^+/\phi^-$.

One of the interesting predictions of the model where the ATIC peak
is explained as consequences of propagation effects from local SNRs,
is that the electron spectrum around the ATIC peak is dominated by
nearby sources. These source produce only primary electrons and have
only negligible contribution to secondary positron flux. As a result
ATIC observations force the electron/positron ratio to start
decreasing at a few hundred GeV, which is not far above the present
PAMELA measurement. It should reach a minimum around the ATIC peak,
where it should start rising again. Whether or not it can go up to
about $50$\% at a few TeV depends on whether the CRs from very
recent SNe, the Cygnus Loop and Vela, could have reached us or not.
This critically depends on the exact diffusion coefficient.  Here it
is also worth pointing out that above a few TeV the secondaries must
be produced within the local bubble, implying that their
normalization should be ten times lower than for the lower energy
secondaries. These predictions are in contrast to the case where the
ATIC peak is due to a primary source of pairs, in which case the
positron fraction is expected to keep rising also at a few hundreds
GeV. With these predictions, it will be straightforward in the
future to distinguish between propagation induced ``anomalies", and
real anomalies arising from primary pairs (in particular, when
PAMELA's observations will extend to higher energies). Of course, it
is possible that the ATIC peak is due to a source of primary pairs,
while the PAMELA anomaly is a result of SNRs in the spiral arms, but
then it would force us to abandon the simplicity of the model, that
the anomalies are all due to propagation effects from a source
distribution borne from the known structure of the Milky Way.

Irrespectively, this work demonstrates that the intermediate scale
inhomogeneities expected in the CR source distribution leave
nontrivial imprints on the electron and positron spectra. These
should be further investigated before reaching definitive
conclusions about the existence of primary positron sources.


We thank Marc Kamionkowski, Re'em Sari and Vasiliki Pavlidou for
helpful discussions. The work was partially supported by the ISF
center for High Energy Astrophysics, an ISF grant (NJS), an IRG
grant (EN) an ERC excellence grant and the Schwartzman Chair (TP).

\def\apj{Astrophys.\ J.}
\def\nat{Nature}
\def\apjl{Astrophys.\ J. Lett.}
\def\aap{Astron.\ Astrophys.}
\def\prd{Phys. Rev. D}
\def\physrep{Phys.\ Rep.}



\end{document}